\documentclass[10pt,conference]{IEEEtran} 
\usepackage[utf8]{inputenc}
\usepackage[square,sort,comma,numbers]{natbib}
\title{Towards the quality improvement of cross-platform \\mobile applications}
\usepackage{balance}
\usepackage{natbib}
\usepackage[bookmarks=false]{hyperref}
\usepackage{natbib}
\usepackage{graphicx}
\usepackage{color}
\usepackage{paralist}

\newcommand{\xpmf}{{cross-platform mobile app development frameworks}}
\newcommand{\xpmfes}{{cross-platform mobile app development frameworks }}
\usepackage[most]{tcolorbox}
\usepackage{verbatim}
\usepackage{lipsum}
\usepackage[framemethod=default]{mdframed}
\newmdenv[linecolor=black,backgroundcolor=gray!30]{myframe}
\newcommand{\negrita}[1]{{\bf{#1}}}

\author{\IEEEauthorblockN{
Matias Martinez, Sylvain Lecomte
}
\IEEEauthorblockA{
Univ Lille Nord de France, F-59000 Lille, France,\\
       Universit\'e de Valenciennes et du Hainaut Cambr\'esis, \\Laboratoire LAMIH - UMR CNRS 8201\\
59313 Valenciennes Cedex 9\\
nom.prenom@univ-valenciennes.fr
}
}

\begin{document}

\maketitle

\begin{abstract}
During last ten years, the number of smartphones and mobile applications has been constantly growing. Android, iOS and Windows Mobile are three mobile platforms that cover almost all smartphones in the world in 2017. 
Developing a mobile app involves first to choose the platforms the app will run, and then to develop specific solutions (i.e.,  native apps) for each chosen platform using platform-related toolkits such as Android SDK.  
A \emph{cross-platform mobile application} is an app that runs on two or more mobile platforms. 
Several frameworks have been proposed to simplify the development of cross-platform mobile applications and to reduce development and maintenance costs. 
They are called {\it \xpmf}.
However, to our knowledge,  the life-cycle and the quality of cross-platforms mobile applications built using those frameworks have not been studied in depth.  
Our main goal is to first study the processes of development and maintenance of mobile applications built using \xpmf,  focusing particularly on the bug-fixing activity. 
Then, we aim at defining tools for automated repairing bugs from cross-platform mobile applications.
\end{abstract}

\section{Introduction}



\subsection{Context}

Studies estimate that currently in 2017, there are 2 billions smartphones in the world\footnote{\url{https://www.statista.com/statistics/330695/number-of-smartphone-users-worldwide/}} 
and this number will increase 50\% the next 5 years. Smartphones are mobile devices that run software applications such as games, social network (Facebook, Twitter) and banking apps. A \emph{native mobile application} is an app built to run in a particular mobile platform. 
Currently, there are three platforms that dominate the smartphone market: Android (from Google), iOS (from Apple) and Windows Mobile (from Microsoft). A large number of applications that run on those platforms are available in application stores. For instance, Google Play, the official app store of Android applications,  has more than 2 millions apps available to download.

A \emph{cross-platform mobile application} is an application that targets more than one mobile platform. 
A manner of building this kind of apps is, for example, to develop a \emph{native} application created for running on Android devices (built using Java technology) and another \emph{native} application built (using Objective-C language) for iOS devices. 

A current challenge for business enterprises, software companies, and independent mobile developers is to choose the target platforms for their mobile applications. 
To cover a large number of users, companies and developers aim at releasing their mobile apps on the three mentioned platforms. More platforms targeted involves to potentially reach more users and, thus, to increase the impact on the market.

However, nowadays, creating a cross-platform application as we mentioned, implies the development of one native application for each platform to target. Thus, companies have to afford the cost of two or  more development processes of their cross-platform apps. It has a monetary impact and also affects the project organization: e.g., a company must find specialized developers for targeting each platform, the development duration for each platform could be different and it could exist technical limitations and differences between platforms, such as permission policies.

\subsection{Cross-platform mobile app development\\ frameworks}

During last years, researches and industry sectors have both focused on proposing solutions for developing mobile applications to overcome the mentioned problematic. 
One of the main objectives of those solutions is to reduce the cost and time of developing and maintaining mobile applications.

From the academic research community, Perchat et al. \cite{perchat2013component,perchat2014common} proposed a framework that allows the integration of cross-platform components in any application (iOS, Android), with the goal of soften the difference between each mobile platform.
Le Goaer et al. proposed Xmob \cite{le2013yet}, a technology-neutral domain-specific language (DSL) intended to be cross-compiled to produce native code for a variety of platforms.

From the industry side, Phonegap (a.k.a. Apache Cordova)\footnote{\url{https://cordova.apache.org}} from Adobe is a frameworks to build mobile applications using CSS3, HTML5, and JavaScript source code. The code is embedded on a skeleton native app, which interprets and executes the app code on the specific device.  This kind of applications are called \emph{Hybrid mobile applications} due they are conformed by both native (e.g., Java for the Android skeleton) and not-native (e.g., HTML) code.

Another cross-platform mobile app development frameworks is Xamarin\footnote{\url{https://www.xamarin.com}}. It makes it possible to do native Android, iOS and Windows development in C\#. 
Developers re-use their existing C\# code, and share significant code across device platforms. For example,  Evolve\footnote{\url{https://github.com/xamarinhq/app-evolve}} a conference event management app, was developed using Xamarin and is around 15,000 lines of code: the iOS version contains 93\% shared code i.e., C\# (the remain 7\% iOS specific code) and the Android version contains 90\% shared code.\footnote{https://goo.gl/eqj1ml} 
Sharing code means to reduce both develop and maintenance cost. 
Frameworks as Xamarin or React-Native\footnote{\url{https://facebook.github.io/react-native}} are know as \emph{Cross-compiled approaches}: their receive as input an application written in a particular not-native programming language and transform it to a native code for a particular mobile platform.

In this section we have listed some \xpmf, but mobile developers can find many more options according to their needs.

\subsection{Developing cross-platform mobile apps}
Cross-platform development frameworks are built for saving development effort and money, and for making more agile the development process by having a single team building the whole solution.  
Researches have conducted different studies about cross-platform development frameworks such as comparison between them \cite{heitkotter2012evaluating,francese2013supporting,dalmasso2013survey,palmieri2012comparison,desruelle2012challenges} or with web based mobile apps \cite{charland2011mobile}. 
However, to our knowledge, no previous studies have investigated the problematic and challenges of \emph{maintain} cross-platform mobile apps built using those frameworks.  

\subsection{Our research objectives}

Our main goal is to study the quality of cross-platform mobile apps built using \emph{cross-compiled frameworks} such as Xamarin or React-Native. 
We envision studying how developers build and maintain cross-platform mobile applications built using those frameworks, and how bugs are fixed during the application life-cycle.
Moreover, we aim at defining a new generation of automated repair approaches for repairing bugs on cross-platform applications. 

To our knowledge, this research  would be the first at studying and understanding the development, maintenance and automated repair processes of cross-platforms mobile applications. 
We consider our research will have a strong impact on the industry of software development, specially for mobile development companies which base their business model on the development of cross-platform apps.

\section{Case Study: History of Facebook mobile application}

In this section, we present a case study that summarizes the different strategies adopted by Facebook developers to build the Facebook mobile apps.

In 2009, the engineer team of Facebook has built the first mobile application of Facebook using an hybrid approach: 
a web-based application (i.e., HTML5 technologies) wrapped on native platforms code such as iOS and Android.
%
By those days, the Facebooks engineers highlighted that this development approach let them leverage much of the same code for iOS, Android, and the mobile web. Furthermore, the approach also allowed them to iterate on experiences quickly by launching and testing new features without having to release new versions of the apps \cite{dannFacebook}.
However, over time, the Facebook engineers realized that users had more expectations in terms of speed and user experience \cite{konicekFacebook}.
For those reasons, they decided to re-write from scratch the native Facebook app for iOS, and later the native app for Android platform. 
Unfortunately, the engineers rapidly realized that 
recompiling on every code change was slow, and separate sets of skills were needed to build for iOS and Android, having as consequence a slower product development and an increase of development costs \cite{dannFacebook}.

In 2015, the adoption of Reach Native framework brings the engineer team the possibility to overcome those limitations.
React Native is a \emph{learn-once-run-everywhere} app development tool, started as a hackathon project in 2013.
In a nutshell, React-Native allows developers to write components in JavaScript language, which are converted to native components for iOS and Android platforms.
Nowadays, Facebook, Instagram and Airbnb are examples of cross-platform mobile apps developed using React-Native\footnote{\url{https://facebook.github.io/react-native/showcase.html}}.

This brief story shows that the development of cross-platform mobile apps is a challenge, even for one of the most valuable companies in the world.

\section{Long-term Research Objectives}
\label{sec:objectives}

In this section we propose three objectives for understanding and improving quality of mobile applications.
We focus on studying application developed using cross-platform mobile development frameworks, particular on Xamarin and React-Native, maintained by Microsoft and Facebook, respectively.
We choose those two frameworks due to  they are open-source and well-documented (manuals, forums and white-papers).

\subsection{Studying development of cross-platform apps}

First, we plan to compare the processes of developing cross-platforms using \emph{Cross-compiled development frameworks} with respect to the development of native apps using tools for that propose (Xcode, Android Studio).
For instance, we plan to study the learning curve of those frameworks. 
Then, we plan to study how a framework manages the differences between platforms and how those differences impact on the app development and on the app's capabilities.  
Finally, we plan to study whether the use of those frameworks helps, as it is expected, to have faster time-to-market, increase component reuse, decrease cost, etc.
Those research goals focus on the {\it development phase} of applications. 
The next two manly focus on the {\it maintenance phase}.

\subsection{Studying the maintenance of cross-platform applications}

The second goal is to study the maintenance of cross-platform applications.
The main research question we plan to develop is: {\it{“Are apps generated using those frameworks easier to maintain that those generated using traditional development of native apps?".}}
In particular, we are interested in studying:
\begin{inparaenum}[a)]
\item what are the most frequent bugs that appear in those cross-platforms apps, 
\item how bugs look like, 
\item how bugs are fixed.
\end{inparaenum}

\subsection{Repairing bugs on cross-platform apps}

Third, we are interested in automated repair those bugs. Based on the knowledge from studying development and maintenance phases of cross-platforms apps, we will focus on approaches that helps developers to overcome the problems emerged from their apps. 
We envision presenting to developers suggestions and candidate patches to fix bugs at the level of source code. 
In case of Xamarin, which most parts of the code are developed in C\#, we aim at defining an approach that generates C\# code repairs.

\section{Proposed experiments}
\label{sec:Experiments}

In this section we discuss how we plan to carried our the experiments for targeting each research goal mentioned in Section \ref{sec:objectives}.

\subsection{Creating a mobile apps corpus}

We plan to study open-source apps mobile application to understand and characterize the development and maintenance of cross-platform mobile applications. We plan to create a corpus of mobile apps developed using  the frameworks we target (i.e., Xamarin and React-Native). 
For that, we plan to explore and mine code repositories platforms such as Github as we have previously done for understanding the bugfixing activity of Java application \cite{martinez2013AST, martinez2013Mining}.
Hence, the main challenge is to identify projects built using \xpmfes from other kind of projects, including native mobile apps. 

For mining projects, we plan to build a tool for retrieving candidates mobile applications.
This tools will carry out two steps: in the first one it queries the hub sites such as Github
for retrieving projects from their meta-data (e.g., project name, description, language).
For example, using the query:
\begin{myframe}
\url{https://github.com/search?q=xamarin+in%3Areadme%2Cdescription+language%3Ac%23&type=Repositories}\end{myframe}
we retrieve from Github candidate Xamarin projects that:
\begin{inparaenum}[a)]
\item are written in C\# (language used to build Xamarin apps);
\item contains the word `Xamarin' in the project descriptions or in the README.md file. 
\end{inparaenum}

In the second step, the tool will parse and analyze the source code of each retrieved project to confirm if it is a mobile app built using the development platform we target (in our example Xamarin).
The former step is important for removing false positives. 
For example, the previous query returns more than 8800 projects.
After manually analyzing the results we concluded that there are two kind of projects we are not interested in analyzing.
First, some projects are plug-ins and sub-components of the development platforms, 
for example, \emph{Forms Labs}\footnote{\url{https://github.com/XLabs/Xamarin-Forms-Labs}} project provides a set of new controls for Xamarin.
Secondly, we found `toy' projects done for testing the development platforms\footnote{Example: \href{https://github.com/TheRealAdamKemp/Xamarin.Forms-Tests}{https://github.com/TheRealAdamKemp/\\Xamarin.Forms-Tests}} and other projects with examples\footnote{Example: \url{https://github.com/conceptdev/xamarin-samples}}.
Those projects have usually few lines of code and do not represent real mobile apps. 
As consequence, for implementing the tool for mining cross-platform mobile projects, we need to 
\begin{inparaenum}[a)]
\item define more precise and complex queries (using, for instance, the Github API\footnote{\url{https://developer.github.com/v3/}});
\item define heuristics for filtering not desired projects (plug-ins, examples, etc.).
\end{inparaenum}

\subsection{Studying bugs and fixes of mobile apps}

Our goal is to create a taxonomy of frequent bugs from mobile apps developed by our target developers frameworks. For doing that, we plan to mining bug fixes and related information from: 
\begin{inparaenum}[a)]
\item issues from report systems such as Bugzilla, Github issues, etc;
\item bug fix commits from our mobile apps corpus;
\item questions and responses posed on forums such as StackOverflow%
\end{inparaenum}.
Once we identify bugs, we plan to create a bug benchmark to be used for evaluating automated program repair approaches.

Our intuition is there are two kinds of bugs on mobile apps developed using \xpmf.
First, a mobile app could fail due to a bug in its code. 
For example, derefrencing a null pointer, an infinite loop, or a wrong method call\footnote{Example: \url{https://goo.gl/1WDnC7}}. 
Secondly, a mobile app could fail due to a bug presents inside the \xpmf used to built the app. 
For example, the bug 45311 on Xamarin Android project\footnote{\href{https://bugzilla.xamarin.com/show\_bug.cgi?id=45311}{https://bugzilla.xamarin.com/show\_bug.cgi?id=45311}} is a common bug in Java: given a class field name $javaUrl$, and a formal argument from a method also called $javaUrl$, then the assignment $javaUrl = javaUrl;$ does not affect the field. The fix consisted on adding the $this$ keyword on the left-part of the buggy assignment\footnote{\href{https://goo.gl/p0F9cq}{https://goo.gl/p0F9cq}}.

The complexity and constant evolution of mobile platforms such as Android or iOS, and the large diversity of mobile devices in the wild involve that  \xpmfes  must evolve at the same speed, involving the potential creation of new bugs. 
In the case of React-Native, which is open-source, Facebook has a development team with around 20 developers \cite{konicekFacebook} for evolving the framework (which includes the addition of new mobile platforms, new features, and bugfixing). 
Surprisingly, the external community also participate on the evolution of the framework: 
in February 2016,  more than 50\% of the commits came from external contributors (not employers of Facebook).

The study of bugs will allow us to identify the most frequent bugs present on mobile apps for each development platform and to define a taxonomy of defect class.

\subsection{Automated repair bugs of mobile apps}

Recent years, researchers have proposed approaches for automatically repair bugs such as \cite{Goues2012journal,nguyen2013,nopol2016}. 
\emph{Test-based repair repair approaches} \cite{Goues2012journal} use test suite as correctness oracles, i.e., they are used as proxies of the program specification. 
In a nutshell, those approaches receive as input a buggy program $p$ and its test suite $ts$, which has, at least, one failing test case (due to the presence of the bug). The approaches produce, when it is possible, a modified version of $p$ which passes all tests from the suite $ts$.
Each repair approach applies a set of \emph{repair actions}, which are source code changes done to create a candidate repair.

Our goal is to define a platform for repairing functional bugs from mobile apps built with Xamarin. 
The platform will re-implement state-of-the-art repair techniques.
For instance, to re-implement the approach GenProg \cite{Goues2012journal} (which the original implementation targets C code) we need:
\begin{inparaenum}[1)]
\item a tool for parsing source code of the buggy mobile app;
\item a tool for manipulating and transforming the parsed code for synthesizing the candidate repair;
\item a \emph{correctness oracle} (e.g., test suites) which decides if a candidate repair is correct or incorrect.
\end{inparaenum}

We have already implemented GenProg for repairing Java code \cite{astor2016}. There, we used the tool Spoon \cite{spoon}  for parsing and manipulating source code, and JUnit framework\footnote{http://junit.org/junit4/} for running the test suite from the program to repair.
To implement a version of GenProg for repairing buggy mobile apps built with Xamarin, we could use, for instance, Roslyn\footnote{https://github.com/dotnet/roslyn}, an open-source C\# compilers with rich code analysis APIs.
Then, for validating correctness of candidate repairs, as for all \emph{test-based repair approaches}, the repair approach needs  a test-suite related to the buggy mobile application.  
Xamarin provides a testing framework called  Xamarin.UITest\footnote{https://developer.xamarin.com/guides/testcloud/uitest/} that allows programmers to write and execute tests in C\# and NUnit for validating the functionality of iOS and Android Apps.
So, our implementation would need to execute those test cases for validating candidate repairs.



\section{Related work}

\negrita{Cross-platform mobile app development \\frameworks:}
In addition to open-source development frameworks such as PhoneGap, Xamarin and React-Native, 
academic researchers \cite{perchat2013component,perchat2014common,le2013yet} proposed solutions with the goal of simplify the development of cross-platforms applications.

There are several works that classify, compare and evaluate cross-platform mobile application development tools \cite{heitkotter2012evaluating,francese2013supporting,dalmasso2013survey,palmieri2012comparison,desruelle2012challenges} to build  hybrid mobile  and native apps.  
Our goal is to empirically study the life-cycle of mobile apps built using some of those tools, such as Xamarin or React-Native. 


\negrita{Empirical studies of Cross-platforms mobile apps:}
Ciman et al. \cite{Ciman2016} analyzed the energy consumption of mobile development. 
Their results showed the adoption of cross-platform frameworks as development tools always implies an increase in energy consumption, even if the final application is a real native application.

\negrita{Hybrid mobile apps:}
Malavolta et al.  \cite{Malavolta2015} have focused on analyzing hybrid mobile apps (e.g., those that uses PhoneGap) available on the app store Google Play and their meta-data (i.e., user ranking and reviews). One of their finding is that the average of end users ratings for both hybrid and native apps are similar (3.75 and 3.35 (out of 5), respectively). 


\negrita{Mining repositories:}
Previous works such as \cite{Pan2008, martinez2013Mining} have studied and classified bug fix from open-source project repositories. Our goal is to carry out similar experiments but focusing on repositories of mobile apps.

\negrita{Automated software repair:}
During last ten years several approaches have been proposed to repair C bugs such as \cite{Goues2012journal} or Java bugs such as \cite{nopol2016}. 
The buggy programs that conform evaluation benchmarks are typically libraries or command-line (console) applications. 

\section{Threats to validity}

In this section we list possible threats to our research.
One thread is to not find `meaningful' open-source mobile app projects for creating the evaluation corpus. 
We are interested in mobile apps
\begin{inparaenum}[a)]
\item with an large number of commits to study and characterize how developers build and maintain applications.
\item  that use diverse components and functionalities from the underlying platform. 
\end{inparaenum}
Another thread is the impossibility to find meaningful and frequent  bugs from issue tracking systems, bug reports, commits, wikis, etc.
Finally, state-of-the-art repair approaches need correctness oracles such as test suite for validating candidate patches. A thread facing repairing mobile apps is to not find open-source project containing (good quality) test suite. This issue does actually affect the whole automated repair discipline \cite{martinez2016automatic}.

\section{Conclusion}

The development and maintenance of cross-platform mobile apps have particular challenges, different from those related to desktop or web apps.
{\it Cross-platform mobile app development frameworks} have emerged with the goal of simplify the development of cross-platform mobile apps, reduce the development, maintenance costs and the time-to-market of the apps.
Our research focuses on understand the maintenance process of those apps with the goal of producing tools for improving the quality of mobile applications.

\negrita{Acknowledge:}
The ELSAT2020 project is co-financed by the European Union with the European Regional Development Fund, the French state and the Hauts de France Region Council.

\bibliographystyle{plain}
\balance
\bibliography{references}

\end{document}